\documentclass[conference]{IEEEtran}
\usepackage{cite}
\usepackage{flushend}
\usepackage{amsmath,amssymb,amsfonts,amsthm}
\usepackage{graphicx}
\usepackage{textcomp}

\usepackage[caption=false,font=footnotesize]{subfig}
    
\newtheoremstyle{plainnoparen}
  {}{}{\itshape}{}{\bfseries}{.}{5pt plus 1pt minus 1pt}
  {\thmname{#1}\thmnumber{ #2}\thmnote{. \textnormal{#3}}}
\theoremstyle{plainnoparen}
\newtheorem{Assumption}{Assumption}
\newtheorem{Lemma}{Lemma}

\newtheorem{Proposition}{Proposition}
\newtheorem{Theorem}{Theorem}

\def\BibTeX{{\rm B\kern-.05em{\sc i\kern-.025em b}\kern-.08em
    T\kern-.1667em\lower.7ex\hbox{E}\kern-.125emX}}
\makeatletter
\def\@IEEEproof[#1]{\@IEEEQEDshowtrue\par\noindent\hspace{\IEEEproofindentspace}{\itshape #1. }}
\makeatother
\begin{document}

\title{Re-Planning Budgets under Admission Control in Grant-Free LEO Beam-Hopping Access
}

\author{\IEEEauthorblockN{Joon-Young Park}
\IEEEauthorblockA{\textit{School of Electrical Engineering} \\
\textit{Korea University}\\
Seoul, South Korea \\
megaduke7@korea.ac.kr}
\and
\IEEEauthorblockN{Young-Chai Ko}
\IEEEauthorblockA{\textit{School of Electrical Engineering} \\
\textit{Korea University}\\
Seoul, South Korea \\
koyc@korea.ac.kr}
}

\maketitle

\begin{abstract}
Reordering low-Earth-orbit (LEO) beam-hopping visits can reduce grant-free update delivery even when each region receives the same service time. We derive the largest limit on inter-frame position changes that meets a prescribed delivery-rate loss tolerance when every group is visited once per frame. The guarantee concerns a protected beam's spatially averaged long-run delivery rate relative to optimized periodic service under exogenous re-planning. Fixed admission can favor irregular visits, whereas admission optimized for each revisit interval makes the per-visit reward concave under sufficient conditions on transmission success. Periodic service is then optimal at a fixed mean interval. The shortest and longest allowed intervals determine the exact worst-case loss, attained by feasible alternating visit orders that respect every group's displacement limit. The resulting budget requires no estimate of the revisit-interval distribution and extends to independent stationary renewal updates. In a collision example with 40 groups and 40 resources per slot, increasing the loss tolerance from zero to 1\% raises the limit from 17 to 24 positions. With independent priorities, the latter retains over 90\% of the priority-weighted waiting-cost reduction achieved by unrestricted ordering. Received-power simulations under Rayleigh fading support the predicted loss.
\end{abstract}

\begin{IEEEkeywords}
admission control, beam hopping, grant-free random access, LEO satellite communications, non-terrestrial networks, re-planning, stochastic ordering.
\end{IEEEkeywords}

\section{Introduction}

Low-Earth-orbit (LEO) satellites support connectivity in regions beyond terrestrial coverage. Because satellite payload constraints limit the number of beams that can operate simultaneously, beam hopping serves different coverage areas in turn~\cite{Lei2020,Lin2023,Zhu2024,Feng2024}. The order of these visits may need to change as regional service priorities evolve. We consider priorities assigned by an external mission controller, independently of the protected group's underlying randomness. Advancing one region's visit also changes the timing of service to others. Operators therefore seek to shorten the waiting time from priority assignment to a region's visit while protecting the long-run update-delivery rate of affected regions.

This concern arises in grant-free status reporting, where devices send updates over shared access resources without first obtaining individual transmission grants~\cite{Jia2022,Lee2023,Yang2022}. Changing the visit order changes the time between successive visits, called the revisit interval, even when each region receives the same total service time. With sporadic updates, longer intervals can bring more devices into contention, whereas shorter intervals can leave access resources unused. Admission control can reduce contention by limiting transmission attempts, but it cannot fill opportunities left unused because too few devices have updates. Thus, preserving the service allocation does not by itself protect the update delivery rate.

We translate a prescribed delivery-rate loss tolerance into a limit on how many positions a group may move between consecutive frames. With fixed admission, the per-visit reward can turn convex at long intervals, so irregular visits can outperform periodic service. Under sufficient conditions on transmission success, admission optimized for each interval restores concavity and yields a limit computable without the revisit-interval distribution.

\begin{figure*}[!t]
\centering
\includegraphics[width=0.8\linewidth]{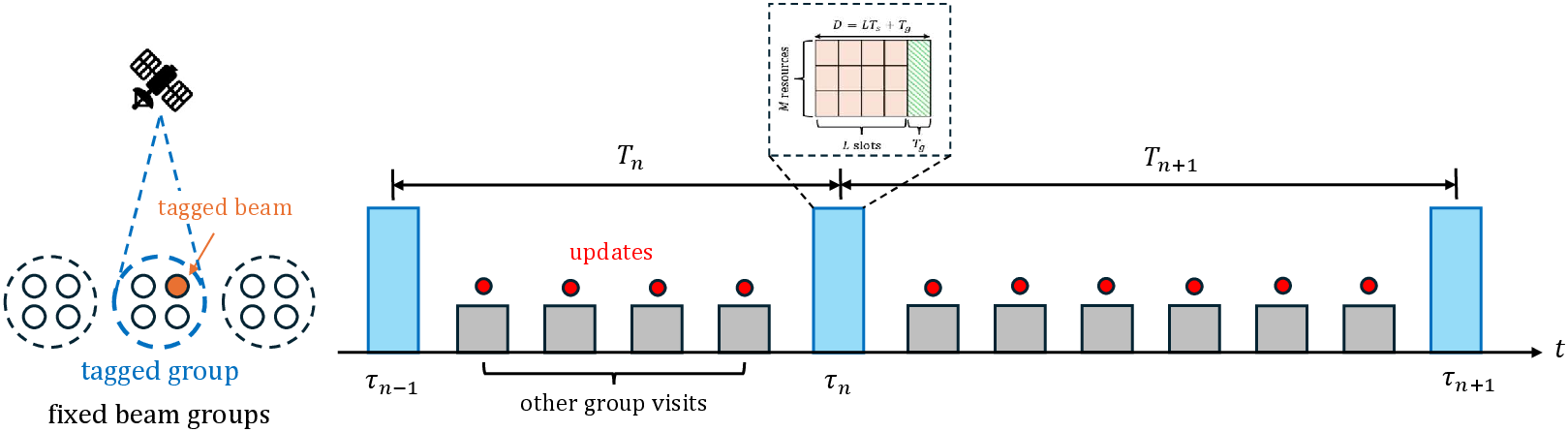}
\caption{Gates $\tau_n$ delimit intervals $T_n$. Each visit has $L$ slots of duration $T_s$ and $M$ resources per slot. Overhead $T_g$ gives visit duration $D=LT_s+T_g$.}
\label{Fig1}
\end{figure*}

Joint access and beam-management designs have been studied in~\cite{Wei2023,Wang2024}. Jeon et al.~\cite{Jeon2026} optimize illumination allocation and hopping patterns for max--min access success under a given activity model. Our question is how far a visit order may be revised, with service allocation fixed, when the resulting revisit intervals also change the probability that a device has an update. Although regular sequences minimize certain queueing costs~\cite{Hajek1985,Altman2003}, this optimality need not carry over to our access model. Age-optimal random access~\cite{Yang2022} targets update age, and beam-hopping queueing analysis~\cite{Feng2024} models packet service. We instead bound service-order changes with fixed per-visit resources. Adaptive access control~\cite{Jia2022} regulates contention, and our revisit-conditioned admission yields concavity and a service-order budget.

For stationary and ergodic revisit sequences independent of the protected group's underlying randomness, our contributions are as follows.
\begin{itemize}
\item We identify how admission control changes the value of revisit regularity. Fixed admission can favor irregular visits, whereas admission optimized for each revisit interval yields a concave reward under sufficient conditions on transmission success. Periodic service is then optimal at a fixed mean interval.

\item We derive an explicit expression for the exact worst-case delivery-rate loss under a limit on every group's position change. Concavity places the worst case at the extreme intervals~\cite{Madansky1959}, and feasible alternating visit orders attain it while respecting all groups' limits. This establishes the largest integer budget meeting a prescribed tolerance within the frame-permutation class.

\item The budget extends to independent stationary renewal updates. Collision- and SINR-law simulations quantify its delivery-rate protection, while priority-driven scheduling experiments measure the waiting-cost reduction it retains.
\end{itemize}

\section{System Model}\label{Sec2}

\subsection{Beam-Hopping Service and Stochastic Revisits}\label{Sec21}
A fixed codebook partitions $B$ beams into $G=B/K$ disjoint groups of $K$ co-active beams, illuminated one group at a time as in Fig.~\ref{Fig1}. Re-planning changes only their order. Each visit has $L$ random-access slots of duration $T_s$ and $M$ orthogonal resources per slot. With switching and control overhead $T_g$, its duration is $D\triangleq LT_s+T_g$. To isolate timing effects, we fix $L$ and $D$ and assume quasi-static footprints and large-scale link statistics over the scheduling timescale.

We evaluate the delivery-rate guarantee for a selected \emph{tagged} beam. Visit $n$ to its group starts at a gate time $\tau_n$, which fixes the revisit interval $T_n\triangleq \tau_n-\tau_{n-1}$ and determines which devices are eligible for that visit. Beams in the group share device density, update rate, and the probability of allowing an eligible device to transmit, termed the admission probability.

To keep a common delivery model across admissible schedules, we exclude feedback from tagged-group updates or access outcomes to gate times. Scheduler information $\mathcal G$ may include other groups' traffic, priorities, tagged-group long-run parameters, and past gate times. Tagged-group randomness $\mathcal F^{\mathrm{tag}}$ contains locations, updates, underlying admission and resource-selection draws, and fading.

\begin{Assumption}[Exogenous re-planning]\label{Ass1}
$\mathcal F^{\mathrm{tag}}$ is independent of $\mathcal G$, and each gate time is determined by scheduler information available up to that time.
\end{Assumption}

To prevent overlapping visits, $T_n\geq T_{\min}\geq D>0$. For long-run averaging, we assume stationary-ergodic, possibly correlated revisits with finite mean $\mu\triangleq\mathbb E[T_n]$, giving time share $D/\mu$. For the displacement budget, we consider frame permutations that visit every group once over $GD$ seconds, so $\mu=GD$. If the tagged group occupies position $J_n\in\{1,\ldots,G\}$ in frame $n$, then $T_n=D(G+J_n-J_{n-1})$. This converts a limit on position changes into a bound on revisit intervals.

\subsection{Latest-Update Traffic and Load-Domain Access Model}\label{Sec22}
Device locations form independent Poisson point processes of intensity $\lambda>0$ in beam footprints of area $A$, with mean population $\bar N=\lambda A$ per beam. Each device independently generates Poisson updates at rate $\xi>0$. At gate $\tau_n$, a device is eligible if it generated at least one update during $(\tau_{n-1}, \tau_n]$, and retains only the latest. Later arrivals wait for the next visit, while undelivered eligible packets expire after this visit. This models status reports whose validity ends with their next scheduled access opportunity. Here, reports not admitted or not decoded are discarded at the end of that visit. Since gate times are independent of tagged-group updates under Assumption~\ref{Ass1}, the eligibility probability given $T_n=t$ is $q(t)\triangleq 1-e^{-\xi t}$.

To control contention, each eligible device is admitted to transmit with probability $p\in[0,1]$, independently across devices. The gateway broadcasts this admission probability before access, with overhead included in $T_g$. Each admitted device transmits once on a uniformly chosen slot-resource pair. We use a device's transmission probability in a given slot as the normalized load $x(t,p)\triangleq q(t)p/L\leq1/L$. Uniform resource selection then gives spatial transmitter intensity $\lambda x(t,p)/M$ per slot-resource pair.

Co-active beams share this load, so all interfering transmitter intensities scale with $x$. With the remaining system parameters fixed, $(t,p,L)$ affect transmission success only through $x$. Let $S(x)$ denote the success probability of an admitted tagged-beam transmission, averaged over device locations, fading, and intra- and inter-beam contention. To compute the mean deliveries per visit, we first denote the mean deliveries per device per slot as $\phi(x)\triangleq xS(x)$. The resulting spatially averaged per-visit delivery count, or reward, is obtained as
\begin{align}
R(t,p)&=\bar Nq(t)p\,S\!\left(\frac{q(t)p}{L}\right)
=\bar NL\,\phi\!\left(\frac{q(t)p}{L}\right).
\label{eq:reward}
\end{align}

\subsection{Admission Classes and Spatially Averaged Delivery Rate}\label{Sec23}
We compare two admission classes. Static admission keeps the same probability across visits, $p_n\equiv\bar p\in[0,1]$. Revisit-conditioned admission uses the interval observed at the gate to set $p_n=p(T_n)\in[0,1]$, with $p$ measurable. Let $F_T$ denote the common distribution of the stationary revisit intervals, and write $T\sim F_T$. Under periodic service, every interval equals $\mu$, so this distribution is the point mass $\Delta_\mu$.

For each scheduler trajectory, we average delivery counts over tagged-group randomness, including device locations. Let $Y_n$ denote the number of deliveries in visit $n$. Dividing accumulated expected deliveries by elapsed time defines the spatially averaged long-run rate
\begin{align}
\bar U_{\rm sp}[p,F_T]
\triangleq
\lim_{V\to\infty}
\frac{\sum_{n=1}^{V}
\mathbb{E}[Y_n\mid\mathcal G]}
{\sum_{n=1}^{V}T_n}.
\label{eq:Usp_def}
\end{align}
A fixed deployment need not attain this ensemble rate. Under Assumption~\ref{Ass1}, however, the rate reduces to a single expectation over $F_T$.

\begin{Lemma}[Delivery-rate representation]\label{Lemma1}
Under Assumption~\ref{Ass1}, stationary-ergodic revisits with $0<\mu<\infty$ satisfy
\begin{align}
\bar U_{\rm sp}[p,F_T]
=
\frac{\mathbb{E}[R(T,p(T))]}{\mu},
\label{eq:Usp}
\end{align}
almost surely with respect to the scheduler process.
\end{Lemma}

\begin{IEEEproof}
Assumption~\ref{Ass1} and one-visit expiry give $\mathbb E[Y_n\mid\mathcal G]=R(T_n,p(T_n))$. Since $0\leq R\leq\bar N$ and $\mathbb E[T]=\mu<\infty$, the ergodic theorem gives convergence of the sample means of $R(T_n,p(T_n))$ and $T_n$ to $\mathbb E[R(T,p(T))]$ and $\mu$, respectively. Their ratio gives \eqref{eq:Usp}.
\end{IEEEproof}

For a given revisit law $F_T$, the best rate attainable with a constant admission probability defines the static benchmark $\bar U_{\mathrm{stat}}^\star(F_T)\triangleq\max_{0\leq\bar p\leq1}\bar U_\mathrm{sp}[\bar p,F_T]$.

\section{Admission Control and the Value of Revisit Regularity}\label{Sec3}
Let $x_c$ denote the load maximizing $\phi(x)$. Fixed admission can make the reward convex at long intervals, whereas admission optimized for each interval makes it concave, which the budget in Section~\ref{Sec34} requires.

\begin{Assumption}[Load-domain access law]\label{Ass2}
The success function $S\in C^2([0,\infty))$ is positive and non-increasing, with $\lim_{x\to\infty}xS(x)=0$. The reward $\phi=xS$ has a unique maximizer $x_c>0$, with $\phi'>0$ below $x_c$, $\phi'<0$ above it, $\phi''\leq0$ on $[0,x_c]$, and $\phi''(x_c)<0$.
\end{Assumption}

The following lemma verifies these conditions for two success-law classes, including the spatial mixtures of the signal-to-interference-plus-noise ratio (SINR) model in Section~\ref{Sec4}.

\begin{Lemma}[Sufficient success-law classes]\label{Lemma2}
Under the smoothness, positivity, monotonicity, and tail conditions in Assumption~\ref{Ass2}, log-concavity of $S$ implies the remaining conditions. Let $(w,\kappa)$ be a random pair with $w\in(0,1]$ and $\kappa\in[\underline{\kappa},\bar{\kappa}]$, where the fixed bounds satisfy $0<\underline{\kappa}\leq\bar{\kappa}<2\underline{\kappa}$. The assumption also holds for the exponential mixture
\begin{align}
S(x)=\mathbb{E}\!\left[we^{-\kappa x}\right].
\label{eq:mixture}
\end{align}
\end{Lemma}

\begin{IEEEproof}
To locate the maximum for log-concave $S$, set $h=-S'/S$, giving $\phi'=S(1-xh)$. Then $h,h'\geq0$, with $h>0$ somewhere by the tail condition. Thus $xh(x)$ is unbounded and strictly increasing wherever positive, so it crosses 1 exactly once, at $x_c$. For $x\leq x_c$, $\phi''=S[-2h+xh^2-xh']\leq-S[h+xh']\leq0$, with $\phi''(x_c)\leq-S(x_c)/x_c<0$.

For \eqref{eq:mixture}, $\phi'=\mathbb{E}[we^{-\kappa x}(1-\kappa x)]$ is positive for $x<1/\bar{\kappa}$ and negative for $x>1/\underline{\kappa}$. Since $\bar{\kappa}<2\underline{\kappa}$, $\phi''=\mathbb{E}[w\kappa e^{-\kappa x}(\kappa x -2)]<0$ on $[0, 1/\underline{\kappa}]$. Hence $\phi'$ has a unique zero with $\phi''(x_c)<0$.
\end{IEEEproof}

\subsection{Fixed Admission Can Favor Irregular Revisits}\label{Sec31}
With fixed admission $\bar p$, the load approaches $\bar x\triangleq \bar p/L$ as the revisit interval grows. Define $x(t)\triangleq \bar xq(t)$ and $R_{\bar p}(t)\triangleq R(t,\bar p)$. At a fixed mean interval, concavity of $R_{\bar p}$ gives periodic optimality by Lemma~\ref{Lemma1} and Jensen's inequality. Differentiating $R_{\bar p}$ twice gives
\begin{align}
R_{\bar p}''(t)=\bar N L\xi^2[\bar x-x(t)]\left[(\bar x-x(t))\phi''(x(t))-\phi'(x(t))\right].
\label{eq:fixed_curvature}
\end{align}
If $\bar x\leq x_c$, Assumption~\ref{Ass2} makes the bracket non-positive, giving concavity. Under full admission, $\bar x=1/L$, so the corresponding slot threshold is $L_c\triangleq 1/x_c$, with concavity for $L\geq L_c$. If $\bar x>x_c$, the bracket is negative at $x=x_c$ but tends to $-\phi'(\bar x)>0$ as $t\to\infty$, making $R_{\bar p}$ eventually convex. If $R_{\bar p}''(\mu)>0$ and $\mu>T_{\min}$, using intervals $\mu-\varepsilon$ and $\mu+\varepsilon$ with equal probability improves the fixed-admission rate for sufficiently small $\varepsilon>0$. Longer revisits approach eligibility saturation, while shorter ones reduce overload.

\subsection{Revisit-Conditioned Admission Restores Concavity}\label{Sec32}
The gateway admits all eligible devices when the full-admission load $q(t)/L$ does not exceed $x_c$ and reduces admission otherwise. The following proposition applies access-probability optimization~\cite{Baccelli2006} to each revisit interval.

\begin{Proposition}[Revisit-conditioned admission]\label{prop:adaptive}
For each interval $t$, maximizing $R(t,p)$ over $p\in[0,1]$ gives
\begin{align}
p^\star(t)&=\min\!\left\{1,\frac{Lx_c}{q(t)}\right\},\label{eq:pstar}\\
R^\star(t)&=\bar N L\,\phi\!\left(\min\!\left\{\frac{q(t)}{L},x_c\right\}\right).\label{eq:Rstar}
\end{align}
If $L<L_c$, the optimal load reaches its cap $x_c$ at
\begin{align}
t_c\triangleq-\frac{1}{\xi}\log\!\left(1-\frac{L}{L_c}\right),\label{eq:tc}
\end{align}
with $p^\star(t)=1$ for $t\leq t_c$ and constant $R^\star(t)$ thereafter. If $L\geq L_c$, $p^\star(t)\equiv 1$, and we set $t_c=\infty$.
\end{Proposition}
\begin{IEEEproof}
Admission realizes any load $x\in[0,q(t)/L]$, so unimodality of $\phi$ gives $x^\star(t)=\min\{q(t)/L,x_c\}$ and \eqref{eq:pstar}--\eqref{eq:Rstar}. For $L<L_c$, solving $q(t_c)/L=x_c$ yields \eqref{eq:tc}. For $L\geq L_c$, $q(t)/L<1/L\leq x_c$, so $p^\star(t)=1$.
\end{IEEEproof}

Since $p^\star(t)$ maximizes each visit reward, Lemma~\ref{Lemma1} makes it optimal among revisit-conditioned policies for any given revisit law $F_T$. Define the resulting long-run rate as $\bar U_{\mathrm{sp}}^\star(F_T)\triangleq\bar U_{\mathrm{sp}}[p^\star,F_T]$. We now ask which revisit laws maximize this rate at a fixed mean.

\begin{Theorem}[Periodic optimality under optimized admission]\label{thm:adaptive}
Under Assumption~\ref{Ass2}, $R^\star$ is non-decreasing and concave. It is strictly concave for $t<t_c$, with a constant branch for $t\geq t_c$ if $t_c<\infty$. Consequently, under Assumption~\ref{Ass1}, periodic service maximizes $\bar U_{\mathrm{sp}}^\star$ at a fixed mean $\mu$. Equality holds exactly when $T=\mu$ almost surely or when $t_c<\infty$ and $\Pr(T\geq t_c)=1$.
\end{Theorem}

\begin{IEEEproof}
To express the capped reward, define $\psi(y)=\phi(\min\{y,x_c\})$. Its concave and constant branches meet with slope zero, so $\psi$ is non-decreasing and concave. Since $q(t)/L$ is increasing and strictly concave and $\psi'>0$ below $x_c$, $R^\star(t)=\bar N L\,\psi(q(t)/L)$ is strictly concave below the cap and constant thereafter. Jensen's inequality gives periodic optimality and the stated equality cases.
\end{IEEEproof}

At a fixed mean, convex order~\cite{Shaked2007} formalizes an increase in the spread of revisit intervals. Write $T_1\leq_{\mathrm{cx}}T_2$ if $\mathbb E[f(T_1)]\leq\mathbb E[f(T_2)]$ for every convex $f$ for which both expectations exist. Applying this definition to $-R^\star$ and using Lemma~\ref{Lemma1} gives
\begin{align}
T_1\leq_\mathrm{cx}T_2\quad\Longrightarrow\quad \bar U_\mathrm{sp}^\star({F_{T_1}})\geq \bar U_\mathrm{sp}^\star({F_{T_2}}).
\label{eq:cx_order}
\end{align}
Admission prevents overload after long revisits but cannot compensate for low eligibility after short ones. When all intervals are at least $t_c$, the reward is constant and variability causes no rate loss at a fixed mean. Since static policies are revisit-conditioned policies and $p^\star(\mu)$ is constant under periodic service, periodic service also maximizes the optimized static rate, with the same optimum for both classes.

\subsection{Distribution-Free Re-Planning Budgets}\label{Sec34}
To quantify loss relative to the periodic optimum of Theorem~\ref{thm:adaptive}, define $\ell(F_T)\triangleq1-\bar U_{\rm sp}^\star(F_T)/\bar U_{\rm sp}^\star(\Delta_\mu)$. Limiting every group's position change between consecutive frames to $m$ confines revisit intervals to $[\mu-mD,\mu+mD]$ while preserving their mean $\mu=GD$. We seek the largest budget guaranteeing $\ell(F_T)\leq\delta$ for a prescribed tolerance $\delta\in[0,1)$ without knowing $F_T$. Since $\mu$ is the midpoint, concavity bounds the expected reward below by the average of the two endpoint rewards~\cite{Madansky1959}.

\begin{Theorem}[Tight re-planning budget]\label{thm:budget}
Under Assumptions~\ref{Ass1}--\ref{Ass2}, consider frame permutations with $T_{\min}=D$, $\mu=GD$. If every group's visit position changes by at most $m\in\{0,\ldots,G-1\}$ between consecutive frames, then
\begin{align}
\ell(F_T)&\leq \ell_\mathrm{max}(m)\triangleq 1-\frac{R^{\star}(\mu-mD)+R^{\star}(\mu+mD)}{2R^{\star}(\mu)},\label{Cor2}\\
m_{\delta}&\triangleq\max\{m\in\{0,\ldots,G-1\}\mid\ell_\mathrm{max}(m)\leq \delta\}.\label{eq:mdelta}
\end{align}
The tagged-beam bound is attainable for every $m$, and $m_\delta$ is the largest budget guaranteeing $\ell(F_T)\leq\delta$ throughout this class.
\end{Theorem}

\begin{IEEEproof}
For $m>0$, the Edmundson--Madansky endpoint bound~\cite{Madansky1959} and Lemma~\ref{Lemma1} give \eqref{Cor2}. The case $m=0$ is periodic. Swap the tagged group at position 1 with the group at $1+m$. Alternating the two orders attains both endpoints equally often and respects every group's bound. An independent uniform initial phase makes the revisit sequence stationary and ergodic. Concavity makes $\ell_\mathrm{max}$ non-decreasing. For every admissible $m>m_\delta$, this construction gives $\ell(F_T)=\ell_\mathrm{max}(m)>\delta$, proving maximality.
\end{IEEEproof}

The interval bound $\mu+mD$ does not bound age of information under repeated failures. For $m>0$, zero worst-case loss is equivalent to $\mu-mD\geq t_c$. Hence the largest zero-loss budget is $m_0=\lfloor G-t_c/D\rfloor$ if $t_c\leq\mu$, and $m_0=0$ otherwise.

\subsection{Conservative Calibration and Renewal Updates}\label{Sec35}
When $x_c$ is uncertain but the eligibility probability $q(t)$ is known exactly, a safe load cap $0<a\leq x_c$ prevents overload. Denote $p_a(t)=\min\{1,La/q(t)\}$, giving $R_a(t)=\bar NL\phi(\min\{q(t)/L,a\})$. The fractional reduction in peak reward is $\epsilon_a \triangleq 1-\phi(a)/\phi(x_c)$. A conservative cap can reduce reward even under periodic service. The bound below quantifies this cost together with the re-planning loss, relative to the original periodic optimum.
\begin{figure*}[!t]
    \centering
    \subfloat[\label{Fig2a}]{%
        \includegraphics[width=0.34\textwidth]{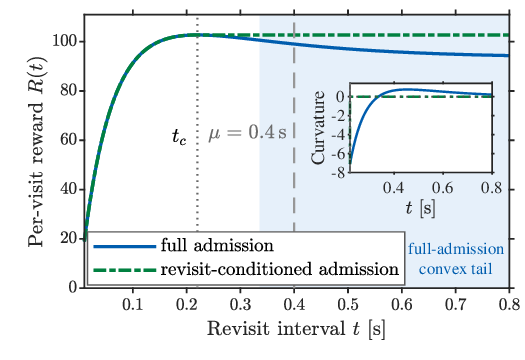}}
    \hspace{0.04\textwidth}%
    \subfloat[\label{Fig2b}]{%
        \includegraphics[width=0.34\textwidth]{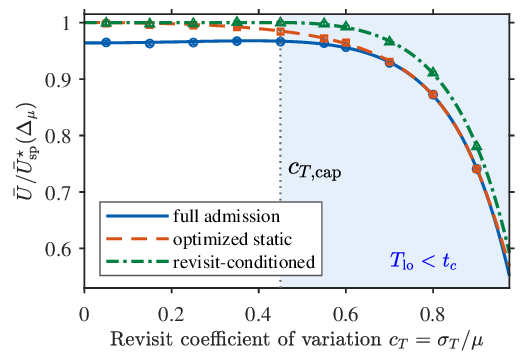}}
    \caption{Admission and revisit variability ($L=8$, $\bar N=450$, $\xi=5$~s$^{-1}$). (a) Per-visit reward; shading marks the full-admission convex tail; inset: normalized curvatures $10^3R''/(\bar N L\xi^2)$ of both policies. (b) Rates normalized by optimized periodic service for $T=\mu(1\mp c_T)$; beyond the dotted line $c_{T,\mathrm{cap}}=1-t_c/\mu$, $\mu(1-c_T)<t_c$ (shaded). Markers: Monte Carlo estimates with pointwise 95\% confidence intervals.}
    \label{Fig2}
\end{figure*}
\begin{figure*}[!t]
    \centering
    \subfloat[\label{Fig3a}]{%
        \includegraphics[width=0.34\textwidth]{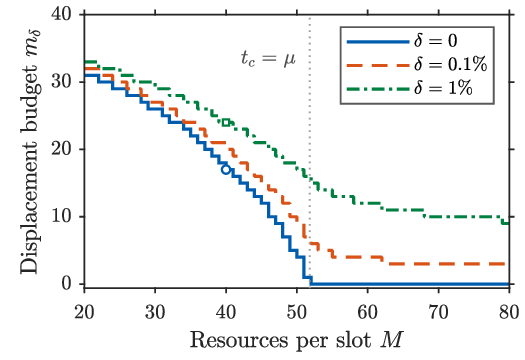}}
    \hspace{0.04\textwidth}%
    \subfloat[\label{Fig3b}]{%
        \includegraphics[width=0.34\textwidth]{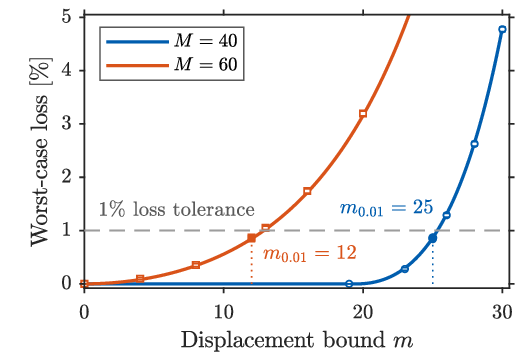}}
    \caption{(a) Collision-law budgets versus resources per slot $M$ at fixed per-resource bandwidth, $D$, and link statistics; dotted line: $t_c=\mu$. (b) SINR-law worst-case loss: $\ell_{\max}$ (lines), received-power estimates with intervals as in Fig.~2 (open markers), and the largest integer budgets meeting the 1\% tolerance (filled markers).}
    \label{Fig3}
\end{figure*}
\begin{Proposition}[Conservative-cap guarantee]\label{prop:conservative}
Under Assumptions~\ref{Ass1}--\ref{Ass2}, $R_a$ is non-decreasing and concave. For the frame-permutation class of Theorem~\ref{thm:budget}, the exact worst total loss is
\begin{align}
\ell_{a,\max}(m)&=1-\frac{R_a(\mu-mD)+R_a(\mu+mD)}{2R^\star(\mu)}\label{eq:conservative}\\
&\leq\epsilon_a+(1-\epsilon_a)\ell_{\max}(m).\nonumber
\end{align}
For the mixture \eqref{eq:mixture}, let $r=\underline\kappa/\bar\kappa$. The safe cap $a=1/\bar\kappa\leq x_c$ gives $\epsilon_a\leq 1-r e^{1-r}$.
\end{Proposition}

\begin{IEEEproof}
Since $a\leq x_c$, $\phi(\min\{y,a\})$ is non-decreasing and concave. Composition with $q/L$ gives the same properties for $R_a$. The endpoint bound and alternating orders of Theorem~\ref{thm:budget} give the exact loss. The pointwise bound $R_a(t)\geq(1-\epsilon_a)R^\star(t)$ follows because the rewards agree when $q(t)/L \leq a$, while otherwise $R_a(t)=\bar NL\phi(a)$ and $R^\star(t)\leq\bar NL\phi(x_c)$. This gives the upper bound.

For the mixture, $\phi'(x)>0$ for $x<1/\bar\kappa$ ensures $a=1/\bar\kappa\leq x_c$; with $u=\kappa/\bar\kappa\in[r,1]$, the bounds $ue^{1-u}\geq re^{1-r}$ and $xe^{-\kappa x}\leq 1/(e\kappa)$ yield $\phi(a)\geq re^{1-r}\mathbb E[w/(e\kappa)]\geq r e^{1-r}\phi(x_c)$.
\end{IEEEproof}
\textit{Renewal updates.} Let devices generate updates by independent stationary renewal processes with inter-update time $X>0$ and $\mathbb{E}[X]=1/\xi$. The equilibrium law~\cite{Ross} gives
\begin{equation}
q_X(t)=\xi\int_0^t \Pr(X>s)\,ds,
\label{eq:qX}
\end{equation}
which is non-decreasing and concave and equals $q(t)$ for Poisson updates. Under Assumption~1, independent thinning at each visit gives (1) with $q_X$ in place of $q$, and the ergodic argument of Lemma~1 applies unchanged; eligibility need not be independent across visits. With $q_X$ in (6)--(7) and $t_c$ solving $q_X(t_c)=Lx_c$ for $L<L_c$, $R^\star$ remains non-decreasing and concave, so (9), Theorem~2, and Proposition~2 carry over; strict concavity and the equality cases of Theorem~1 require separate verification.

\section{Numerical Results}\label{Sec4}
We now evaluate the delivery-rate protection of the budget and the scheduling freedom it leaves. Unless specified otherwise, $G=40$, $K=4$, $L=8$, $M=40$, $\bar N=450$, and $\xi=5~\mathrm{s}^{-1}$. With $T_s=1~\mathrm{ms}$ and $T_g=2~\mathrm{ms}$, $D=10~\mathrm{ms}$ and $\mu=0.4~\mathrm{s}$. Each footprint is a unit disk, and the other co-active beam centers lie four units from the tagged center. At radial distance $d$, the receive pattern is $\gamma(d)=[2J_1(vd)/(vd)]^2$~\cite{TR38811}, where $J_1$ is the first-order Bessel function and $v=1.61634$ gives $\gamma(1)=1/2$. In the collision reference, same-pair interferers independently block reception with probability 1 within the tagged beam and $\beta=0.02188$ across beams; the latter is a blocking proxy based on normalized receive-pattern overlap. Independent decoding probability $S_0=0.93$ gives $S(x)=S_0e^{-\kappa x}$ with $\kappa=(\bar N/M)(1+3\beta)$, so that $x_c=1/\kappa\approx0.08341$ at $M=40$.

Fig.~\ref{Fig2}(a) shows how admission prevents overload after long revisits. Full-admission reward declines after $t_c=0.2201~\mathrm{s}$, whereas revisit-conditioned admission maintains the peak. The full-admission curve becomes convex beyond $0.3356~\mathrm{s}$, as predicted by \eqref{eq:fixed_curvature}. Even periodic full admission loses 3.6\% of the optimized rate because $q(\mu)/L=0.108>x_c$. Fig.~\ref{Fig2}(b) varies the coefficient of variation $c_T=\sqrt{\mathrm{Var}(T)}/\mu$ using equally likely intervals $T=\mu(1\mp c_T)$. For this interval family, irregular visits improve the full-admission rate by up to 0.385\% relative to periodic full admission. Revisit-conditioned admission is lossless until the shorter interval falls below $t_c$, at $c_T\approx0.45$. At $c_T=0.6$, its loss is 0.859\%, versus 3.728\% with optimized static admission. Monte Carlo estimates using $10^6$ independent paired packet trials at each of ten $c_T$ values, with shared draws across endpoints and policies, agree with the predicted normalized rates within 0.319 percentage points. 

The gap between the two admission classes widens with load. With $\bar N=1000$ in every co-active beam and all other parameters unchanged, $t_c$ falls to approximately $0.071~\mathrm{s}$, below both revisit intervals at $c_T=0.8$. Revisit-conditioned admission is therefore lossless by Theorem~\ref{thm:adaptive}, whereas re-optimized static admission loses 12.9\%. A further $10^6$ paired trials agree within 0.16 percentage points. The corresponding budgets are $m_0=32$ and $m_{0.01}=34$.

Although more resources raise the periodic rate, Fig.~\ref{Fig3}(a) shows declining re-planning budgets. Increasing $M$ raises $x_c$, requiring longer intervals to reach the reward plateau. The dotted line marks $t_c=\mu$, beyond which only positive loss tolerances permit reordering. At $M=40$, $m_0=17$ and $m_{0.01}=24$. The latter corresponds to the endpoint distribution with $c_T=m/G=0.6$ in Fig.~\ref{Fig2}(b), whereas $m=25$ incurs 1.261\% loss and violates the tolerance. Fig.~\ref{Fig3}(b) tests the bound under Rayleigh SINR reception. We use independent fading, equal powers, unit boresight mean power, linear threshold $\vartheta=10$, and noise $\nu=0.01$. Let $Z_0$ and $U_j$ be uniform locations in the tagged and $j$th footprints, with $j=0$ tagged. Weighting all signal powers by the tagged receive pattern gives~\eqref{eq:mixture} with $w=e^{-\vartheta\nu/\gamma(\|Z_0\|)}$ and $\kappa=\kappa(Z_0)$, where
\begin{align}
\kappa(z)=\frac{\bar N}{M}\sum_{j=0}^{K-1}\mathbb E\!\left[\frac{\vartheta\gamma(\|U_j\|)}{\gamma(\|z\|)+\vartheta\gamma(\|U_j\|)}\right].
\end{align}
The mixture satisfies Lemma~\ref{Lemma2}, with $\bar\kappa/\underline\kappa\approx1.18<2$. The 1\% line gives budgets 25 and 12 for $M=40$ and $60$, respectively. Markers from $4\times10^6$ independent tagged-transmission trials per $M$ agree with the predicted loss within 0.036 percentage points. Shared endpoint and periodic draws give paired confidence intervals. Other fading models, including shadowed-Rician fading, require verifying Assumption~\ref{Ass2}. 

Table~\ref{tab:operational} evaluates how much scheduling gain the budget retains. Group $g$ at position $J_{g,n}$ waits $D(J_{g,n}-1)$ from the start of frame $n$. With group priorities $w_{g,n}$, we minimize the priority-weighted waiting cost $C_n=\sum_g w_{g,n}D(J_{g,n}-1)$ by solving a linear assignment problem in which assignments violating a group's displacement limit receive prohibitive cost. Priorities have unit-mean exponential or lognormal marginals, with log-standard deviation 1.5 for the latter. The two distributions share Gaussian first-order autoregressive drivers with persistence $\alpha$, independent across groups and of access randomness. Each distribution and budget maintains its own visit order. Groups use the reception model and $M$ indicated in the table, with admission given by~\eqref{eq:pstar}.

Let $\bar C_m$ be the mean frame cost under budget $m$, with $m=0$ fixing the order and $m=39$ allowing unrestricted reordering. The retained fraction of cost reduction is $\eta_m=(\bar C_0-\bar C_m)/(\bar C_0-\bar C_{39})$. We evaluate pooled delivery-rate loss across all groups using the periodic reference $R^\star(\mu)$ of Theorem~\ref{thm:adaptive}. For group $g$'s simulated revisit interval $T_n^{(g)}$, the predicted rewards give $\ell_m=1-\langle R^\star(T_n^{(g)})\rangle/R^\star(\mu)$. Brackets average over groups, frames, and 12 seeds, each with 20,000 frames after 2,000 burn-in frames. Pointwise 95\% intervals, from paired seed bootstrapping for $\eta_m$ and Student-$t$ statistics across seed means for $\ell_m$, lie within 0.117 and 0.0044 percentage points of each estimate, respectively. Gain ranges in Table~\ref{tab:operational} span the two priority distributions, whose losses agree at the displayed precision.
\begin{table}[!t]
\centering
\caption{Retained scheduling gain and delivery-rate loss (\%) with independent priorities ($\alpha=0$).}
\label{tab:operational}
\footnotesize
\setlength{\tabcolsep}{4pt}
\begin{tabular}{lrrrr}
\hline
Model & $M$ & $m$ & $\eta_m$ & $\ell_m$\\
\hline
Collision & 40 & 17 & 75.53--80.99 & 0.000\\
Collision & 40 & 24 & 90.73--94.47 & 0.195\\
Collision & 40 & 39 & 100.00       & 1.147\\
\hline
SINR      & 40 & 25 & 92.18--95.52 & 0.170\\
SINR      & 60 & 12 & 58.71--63.54 & 0.541\\
\hline
\end{tabular}
\end{table}

Table~\ref{tab:operational} shows that the 1\% budgets at $M=40$ retain over 90\% of the unrestricted cost reduction. With persistent priorities ($\alpha=0.9, 0.99$), the zero-loss budget $m=17$ retains at least 99.97\%. A small pooled loss does not, however, establish a per-beam guarantee. Evaluating~\eqref{Cor2} at $m=39$ for the alternating construction of Theorem~\ref{thm:budget} gives a 40.8\% tagged-beam loss while the other 38 groups remain periodic. The certified budget excludes this construction.

Population errors are asymmetric. In the reference collision setting at $m=24$ with exact $q$ and $\beta$, a 20\% overestimate yields a worst total loss of 1.64\% by Proposition~\ref{prop:conservative}, including 1.55\% under periodic service. A 20\% underestimate instead places the cap above $x_c$ and breaks concavity: periodic service loses 2.65\%, more than the 2.18\% endpoint loss. For renewal updates with $\mathbb{E}[X]=0.2~\mathrm{s}$, periodic updates with independent uniform phases raise $(m_0,m_{0.01})$ from $(17,24)$ to $(26,29)$.

\section{Conclusion}
In this paper, we derived the largest integer displacement budget meeting a delivery-rate loss tolerance for a protected beam under frame-permutation scheduling. Optimized revisit-conditioned admission provides the required reward concavity, whereas a fixed admission probability can favor irregular visits. In the reference collision setting with $M=40$ and independent priorities, a 1\% loss tolerance retains over 90\% of the cost reduction achieved by unrestricted reordering. Conservative caps bound calibration loss, and the budget extends to independent stationary renewal updates. The guarantee is spatially averaged and requires exogenous re-planning, homogeneous loading, and one-visit expiry. Extending the guarantee to age of information and to backlogged traffic is left for future work.

\end{document}